\documentclass[%
 reprint,
nofootinbib,
 amsmath,amssymb,
 aps,
prd,
]{revtex4-2}

\usepackage[hidelinks,colorlinks=true,linkcolor=blue,citecolor=blue]{hyperref}

\usepackage{natbib}
\usepackage{CJK}
\usepackage{graphicx}
\usepackage{dcolumn}
\usepackage{bm}

\usepackage{orcidlink}

\begin{document}


\title{Limits of vacuum-template subtraction for LISA massive black hole  binary sources \\ in realistic environments}

\author{Lorenz Zwick}
\email{lorenz.zwick@nbi.ku.dk}
\affiliation{Center of Gravity, Niels Bohr Institute, Blegdamsvej 17, 2100 Copenhagen, Denmark.}
 %


%


\date{\today}

\begin{abstract}
We investigate the impact of gravitational wave (GW) dephasing due to gas accretion on the subtraction of massive black hole (MBH) binary signals over 4 yr of LISA data in the context of the global-fit. Based on state of the art predictions for the population of merging MBHs, we show that imperfect subtraction with vacuum waveform templates leaves a GW residual with an SNR of $3.2^{+5.4}_{-1.9}\times \sqrt{f_{\rm Edd} \langle \dot n \rangle/(20\, {\rm yr}^{-1})}$, where $f_{\rm Edd}$ is the typical Eddington ratio and $\langle \dot n \rangle$ the mean merger rate of LISA MBH binaries. We characterize the dependence of the residual on key population hyper-parameters and provide a simple fitting function. {Finally, we consider the distinguishability of the residual as a stochastic signal by comparing with Bayesian power-law sensitivity curves, while discussing several additional detection and mitigation strategies. Overall, our analysis indicates that the residual is unlikely to be confidently distinguishable from instrumental noise while nevertheless being likely to bias the inference of other signals.
}

\end{abstract}

\maketitle

\section{Introduction}
\label{sec:Intro}
A crucial difference between space-based gravitational-wave (GW) observatories and their ground-based counterparts lies in the approach to data analysis: Signals detected by ground-based interferometers are typically short-lived, separated in time and frequency \citep[see e.g.][]{LIGOdetector,2018maggiore}. Instead, sources populating the mHz band probed by the Laser Interferometer Space Antenna\footnote{Among other proposed space based detectors \citep{TianQin,2021tian}.} (LISA) will be composed of more diverse, long-lasting and overlapping GW signals \citep{2017lisa,2019lisa,2024lisa}. As a consequence, LISA data analysis relies on a ``global fit" approach \citep{Bender1997,Cutler1998,2007cornish,2010babak,2023Littenberg,2024lisa,2025deng}, in which many signals must be modeled and subtracted simultaneously rather than sequentially {\citep[see also Refs.][for examples involving the galactic white dwarf binary foreground]{2005edlund,2017robson,2025khuklaev,2025johnson}}. The necessity of a global fit introduces a number of challenges for properly distinguishing between true signals and noise. Several studies have explored the consequences of using simplified or imperfect GW templates when fitting for simultaneous signals, showing that modeling inaccuracies can lead to imperfect source subtraction and the appearance of un-physical biases or residual features in the data \citep{2005cornish,2007cutlervallis,2014chatz,2021antonelli,2025katz}.

As a further complication, the vast majority of these works are relying on the assumption that both the true signals and the templates are fully described by vacuum General Relativity, such that additional physical effects induced by the environments of inspiralling binaries \citep[known as environmental effects, hereafter EEs,][]{1993chakrabarti,1995ryan,2008barausse,2007levin,kocsis,2014barausse,inayoshi2017,2017meiron,2017Bonetti,2019alejandro,2019randall,2020cardoso,DOrazioGWLens:2020, 2022liu,2022xuan,garg2022,2022cole,2022chandramouli,2022sberna,2023zwick,2023Tiede,2024dyson,2021alejandro,2022destounis,2022cardoso,2020caputo,2022zwick,2023aditya,2024zwicknovel,2025Zwick_ecc,2021andrea,2024basu,2024santoro,2025vicente,2025dyson,2025destounis,2025copparoni,2025torres} are neglected. However, this assumption is at odds with the astrophysical formation channels of LISA sources, which instead involve sustained interactions with gaseous or stellar media prior to and during their GW-driven evolution \citep[see][and references therein]{2022lisaastro}. Among these sources, massive black hole (MBH) binaries produce the loudest signals in the LISA band. The evolution of MBH binaries from galactic separations down to the GW–dominated regime proceeds through multiple stages of hardening, mediated by interactions with the surrounding stellar population and/or the presence of a circumbinary gas disc \citep{2005merritt,Begel:Blan:Rees:1980, Milosavljevic:2003,2006sesana,2009haiman,2015capelo,2020souzalima,2022bortolas,2025spadaro}. In particular, it is expected for gas-assisted hardening to a key role in delivering binaries into the LISA band, a fact that is motivating the recent body of work devoted to characterizing MBH binary–disc interactions \citep{tiede2020,2021doraziodisc, Dittmann:2022, Penzlin:2022, Franchini:2022,2023siwek,SBCodeComp:2024,2025dittman}. Numerical simulations suggest that the influence of gas persists throughout the GW-driven inspiral \citep{Farris_dec+2015, Dittmann:Decoupling:2023, Krauth:2023, Avara:3DMHD:2023, Cocchiararo:2024, ONeillTiedeDOrazio:2025}, as both dynamical and viscous processes allow the disc to remain coupled to the binary. Beyond a plethora of interesting electromagnetic counterparts \citep{2018Tang,2022bogdanovic,2022mangiagli,DOrazioCharisi:2023,2024siwek,D'Orazio:binlite:2024,2024franchini}, this residual interaction with gas can also induce environmental effects (EEs) in the GW signal, most notably in the form of a cumulative phase shift with respect to the vacuum evolution  \cite[see e.g., Refs.][]{2014barausse,2023zwick,garg2022,2024garg}). For individual MBH binaries, this environmentally induced dephasing is often marginal, detectable only in a subset of lighter, low-redshift sources. However, any mismatch between the true signal and the fitted template necessarily leaves behind a residual. Even in a regime where such effects are individually undetectable, the collective impact of residuals due to unmodelled EEs in a global-fit framework increases in tandem with the number of simultaneous signals.

In this work, we characterize the limits of using vacuum waveform templates to subtract MBH binary signals from sources that are merging in realistic astrophysical environments, {using circumbinary discs as a representative case}.  We show that accumulated residuals from unmodelled EEs are likely to produce a measurable power excess over the LISA noise power spectral density (PSD), analogous to the stochastic foreground generated by unresolved Galactic white dwarf binaries. We quantify this effect starting from state of the art predictions for populations of merging MBH binaries and realistic prescriptions for gas-induced dephasing {for accreting, comparable mass binaries}, determining how the resulting PSD feature depends on key population parameters. Our results showcase a previously unexplored contribution to LISA noise and emphasize the importance of EEs not only for parameter estimation of individual sources, but also for the global characterization of the LISA data stream.

The paper is structured as follows: In Sec. \ref{sec:Methods}, we describe our choices for parameterizing massive black hole binary populations, as well as outline the relevant GW physics. In Sec. \ref{sec:Results}, we characterize the resulting residuals and quantify their dependence on the parameters of the population distributions for merging binaries. Finally, we discuss the implications of our findings and summarize our conclusions in Sec. \ref{sec:conclusion}.

\section{Methods}
\label{sec:Methods}
\subsection{Massive binary populations and their environments}
\label{sec:Methods:pop}

\begin{figure*}
    \centering
    \includegraphics[width=1\linewidth]{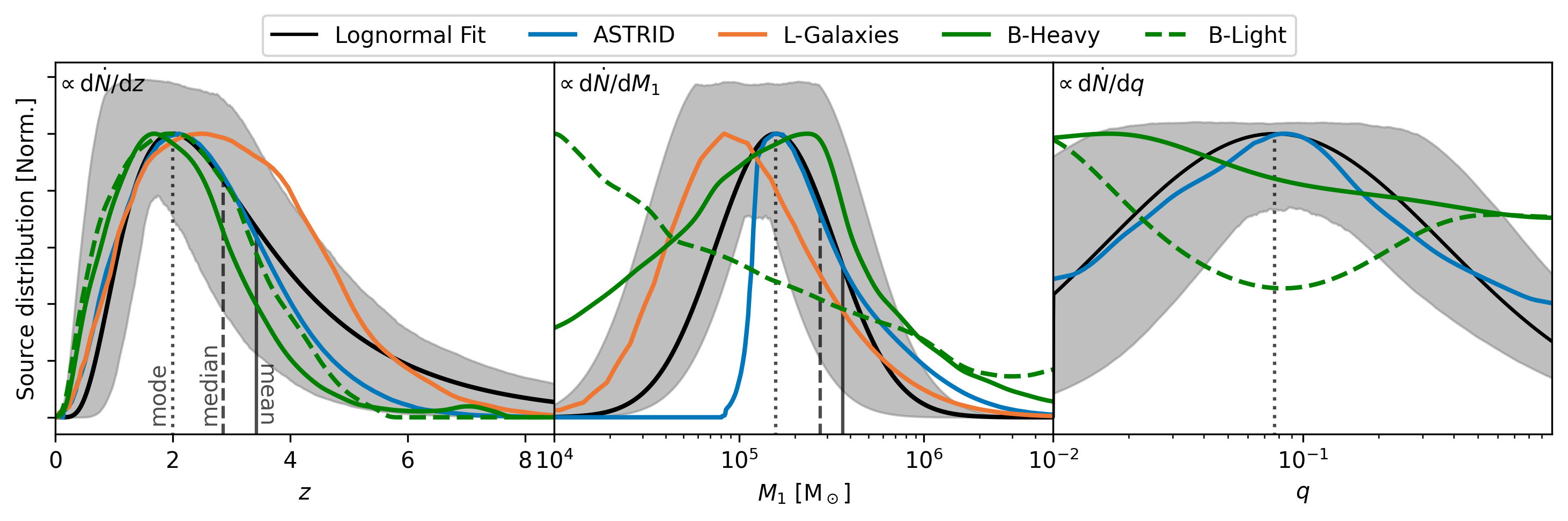}
    \caption{Normalised differential merger rates from the ASTRID simulation \citep[blue][]{2025wang}, L-Galaxies \citep[orange][]{2024izquierdo} and \cite{2020barausse} (green). We employ a lognormal parametrization (black) with representative uncertainties (grey) to broadly reproduce the results of and the variation between different predictions. The distribution in mass ratio $q$ is truncated between 0.01 and 1.}
    \label{fig:pop_fit}
\end{figure*}
The formation and evolution of MBH binaries from galactic scales down to sub-parsec separations is a long-standing problem in astrophysics \citep{Begel:Blan:Rees:1980,Milosavljevic:2003}. Both cosmological simulations and semi-analytical approaches are used to provide quantitative predictions for the merger rate of MBH binaries \citep{2007micic,Volonteri:2012kq,2020volonteri,2022chen}. In practice, such population models yield differential merger rates of the form:
\begin{equation}
\frac{\mathrm{d} \dot{n}}{\mathrm{d}x} \propto \mathcal{P}_{x}(x)
\end{equation}
where $\dot{n}$ is the merger rate, $\mathcal{P}_x$ the probability distribution and $x$ represents one or more source parameters, such as the primary BH mass $M_1$, the binary mass ratio $q< 1$, or the redshift $z$. The detailed physical pathways for MBH binary formation remain uncertain, and differences in the underlying assumptions of simulations and semi-analytical models range from prescriptions for BH seeding \citep{2005sesana,2024reganvolonteri} and accretion \citep{Springel:2005,2023Liao} to the efficiency hardening via gas, stars or additional MBH \citep{blaes2002,2006berczik,2007mayer,2011khan,2016bonetti,2021varisco}. Nevertheless, recent population studies have broadly converged in their predictions for MBH binaries detectable by LISA. Most models predict a merger rate of few $\mathcal{O}(10)$ events per year, with the bulk of the population characterized by primary masses in the range $M_1 \sim 10^{5}\,\rm{M}_\odot$ to $10^{7}\,\rm{M}_\odot$ and typical mass ratios $q\gtrsim 0.1$. The redshift distribution is broad, with mergers occurring predominantly at $z \sim 1$ to $4$, roughly following the galaxy merger rate shifted by a delay \citep[see e.g.][as a general reference]{2024lisa}. 

In this work we parameterize the differential merger rate as a product of independent distributions:
\begin{equation}
\frac{\mathrm{d}^3 \dot{n}}{\mathrm{d}z\,\mathrm{d} M_1\,\mathrm{d} q}
\propto \mathcal{P}_{M_1}(M_1)\, \mathcal{P}_{q}(q)\, \mathcal{P}_{z}(z),
\end{equation}
where each of the marginal distributions $\mathcal{P}_x$ is modeled as a lognormal function. The lognormal in base 10 reads:
\begin{align}
    \mathcal{P}_x \propto \frac{1}{x\,\ln(10)\,\sigma_x\sqrt{2\pi}}
\exp\!\left[
-\frac{\left(\log_{10} x/ x_{\rm c}\right)^2}{2\sigma_x^2}
\right],
\end{align}
where the median $x_{\rm c}$ and widths $\sigma_{x}$ are calibrated by fitting to the results of several recent population studies based on independent modeling frameworks (see Figure \ref{fig:pop_fit}). Rather than selecting a single population model, we define representative uncertainty envelopes that bracket the range of predictions from these simulations. This uncertainty is propagated throughout our analysis to assess the range of possible residual backgrounds against plausible variations in the underlying MBH binary population. Finally, we model the amount of mergers within a 4 year observation time for LISA as a Poisson distribution with mean $\langle N\rangle=4\, {\rm yr}\,\langle \dot n \rangle$, where the brackets denote the mean.

Our parametrization for the population of MBH mergers is shown in Fig. \ref{fig:pop_fit}. We base our fiducial population parameters (hereafter ``hyper-parameters") for the lognormal approximations on results adapted from recent analysis of the ASTRID simulation \citep{2025wang}, recent results from the L-Galaxies semi-analytical suite \citep{2024izquierdo}, as well as the analysis performed in Ref. \cite{2020barausse}, distinguishing between a heavy-seed model and a light-seed model \footnote{For the data adopted from \cite{2024izquierdo} and \cite{2020barausse}, we performed an additional step of transforming the differential rates from total mass ${\rm d}\dot{n}/{\rm d}M_{\rm tot}$ to primary mass. Additionally, we note that differential rates are typically reported in $\log M$, while we work directly with the primary mass.}. These works represent fairly typical estimates for the  differential merger rates rates, and their results are directly accessible from the respective papers. Thus, our fiducial hyper-parameters and uncertainty envelopes are as follows:
\begin{align}
\label{eq:popparam1}
    x_z &= 2.8 ^{+ 1.2}_{- 1.2} \\
    \mu_z &= 0.26^{+ 0.15}_{- 0.15} \\
    x_{M_1} &= 2.8^{+2.7}_{-2.0} \times 10^5\, [\rm{M}_{\odot}] \\
    \mu_{M_1} &= 0.32^{+0.05}_{-0.05}\\
    x_q&= 0.9^{+0.1}_{-0.5}\\ \label{eq:popparam2}
    \mu_{q} &= 0.7^{+0.2}_{-0.2},
\end{align}
where the distribution in $q$ is additionally truncated between $0.01$ and $1$ to avoid unphysical mass ratios and to remain in the MBH range. The fiducial values and uncertainties shown above are heuristically via a combination of least-square fitting and manual adjustments: {First, we take the MBH merger distributions from the respective papers, interpolate them and resample them evenly. Then, we use a simple minimization routine to find the least-squares best fit for each set of distributions, with equal weights across all points. This procedure yields the best fit values and the uncertainties for the parameters of the log-normal distributions. We note here that the precise results of this fitting procedure do not matter, since the uncertainties are large and strongly dominated by the intrinsic variance of the different population models. Rather, this procedure is performed in order to efficiently} sample large ensembles of MBH binaries consistently with current expectations, while also having direct control over the hyper-parameters that most strongly influence the accumulation of residual power in the LISA data stream. This will be crucial to understand the behavior of the residuals when varying population properties.
\newline \newline

As mentioned previously, {all MBH binaries require additional forms of hardening to bridge the gap between galactic and sub–parsec separations. In this work,} we assume that the MBH binaries in our population are driven into the GW regime primarily through gas-assisted hardening \citep[][]{2004escala,2009cuadra,2017Goicovic,2021bortolas}, and are therefore embedded in circumbinary accretion discs. {This particular choice is motivated by the following considerations: Firstly, the fraction of active galaxies increases with redshift, peaking at values of $\sim 10\%$ to 50$\%$ at redshifts of $\sim 1$ to $3$ \citep{2007eastman,2015georgakakis,2017bufanda,2017wang,2024gatica}. This indicates that the presence of gas in galactic centers is ubiquitous at redshifts corresponding to the peak merger rates for LISA MBHs. Secondly, many theoretical and observational works indicate a connection between galaxy mergers, source of the MBH binaries, and the ignition of high accretion phases in galactic nuclei \citep{Barnes_Hernquist_1996,Hopkins_Quataert_2010,Barnes_2002,Capelo_Dotti_2017,Blumenthal_Barnes_2018,2019aird}. Therefore, the baseline provided by the fraction of active galaxies is, most likely, a lower bound for the fraction of MBH binaries facilitated specifically by gas hardening.}

{These same arguments are motivating the large body of work specifically studying the evolution of binaries in circumbinary discs. Over the last years, several simulations have indicated that gas surrounding the binary can remain dynamically coupled to the system even deep into the GW-dominated regime \citep{Farris_dec+2015, Dittmann:Decoupling:2023, Krauth:2023, Avara:3DMHD:2023, Cocchiararo:2024, ONeillTiedeDOrazio:2025}, and that LISA may be sensitive to secular gas-driven modifications to the binary’s orbit \citep{garg2022, 2024garg, 2023Tiede}}. {Across these works,} a commonly used parametrization of the effect of gas is given by the following prescription \citep[see e.g.][]{2021doraziodisc}:
\begin{align}
\label{eq:gaschirp}
    \dot{\Omega}_{\rm gas} = A\frac{1 + q}{q} \frac{\dot{M}}{M_{\rm tot}}\Omega,
\end{align}
where $\Omega$ is the binary orbital frequency, and $A \approx 3$ is a coefficient calibrated to numerical simulations. {This particular parameterization links the torques experienced by the binary to the global accretion rate through the circumbinary disk, and is robust for circular binaries with components of comparable mass. In this work, we will add  Eq. \ref{eq:gaschirp} linearly to the intrinsic frequency chirp of the binary due to GW emission \citep[see Ref.][for the possible presence of additional cross-terms]{2025garg}, using it as a phenomenological model for gas effects \citep[see Refs.][for works that derive more complex GW perturbations for binaries in gas]{2022zwick,2024zwick,2025Zwick_ecc}. We also neglect to model eccentricity evolution, with the justification that circularisation due to GW emission will dominate in the LISA band, and that the bulk population of MBH binaries will have negligible eccentricities. Note however, that much richer observations of EEs are possible for eccentric outliers \citep{2020moore,2024garg,2025zwickecc_pop}.} Thus, the presence of a circumbinary disc modifies the binary’s orbital evolution relative to the vacuum case, leading to a cumulative phase shift in the emitted GW signal \citep[see][for a recent review]{pedo}. {As customary,} we scale the mass accretion rate $\dot{M}$ onto the binary in terms of a fraction $f_{\rm Edd}$ of the Eddington rate \citep{1916eddington}:
\begin{equation}
\dot{M} = f_{\rm Edd}\,\dot{M}_{\rm Edd},
\end{equation}
where $\dot{M}_{\rm Edd}$ is:
\begin{align}
    \dot{M}_{\rm Edd}
= \frac{4\pi G M_{\rm tot} m_p}{\eta \, \sigma_{\rm T\,c}}.
\end{align}
Here $M_{\rm tot}$ is the binary total mass, $m_{\rm p}$ is the proton mass, $\sigma_{\rm T}$ is the Thomson scattering cross section and $\eta$ denotes the radiative efficiency of accretion, which we fix to 0.1 throughout this work.

{A few works provide specific estimates for the accretion rates experienced by typical LISA MBH binaries, though large uncertainty remains. Ref. \cite{2022mangiagli} investigates three population models of LISA MBH binaries, showing that the fraction of highly accreting systems can range from $\sim$1\% to $\sim$80\% depending on the choice. References \cite{2020barausse} and \cite{2023Lops} suggest a range of possible values within $f_{\rm Edd} \in (0.01,1)$. Another few works link the appearance of near or super Eddington accretion phases to gas-rich mergers \citep{2017kelley,2022farrah,2023Liao}, which once again are responsible for the creation of MBH binaries in the first place. Due to this uncertainty, we simply adopt $f_{\rm Edd}$ as a fiducial parameter that describes the typical Eddington ratio of a population of MBH binaries with a merger rate of $\langle \dot n \rangle$ and note that the residuals will scale as $\sqrt{f_{\rm Edd}}$.}

\subsection{Gravitational waves, template subtraction and signal-to-noise ratios}
\label{sec:Methods:Wave}
\begin{figure}
    \centering
    \includegraphics[width=1\linewidth]{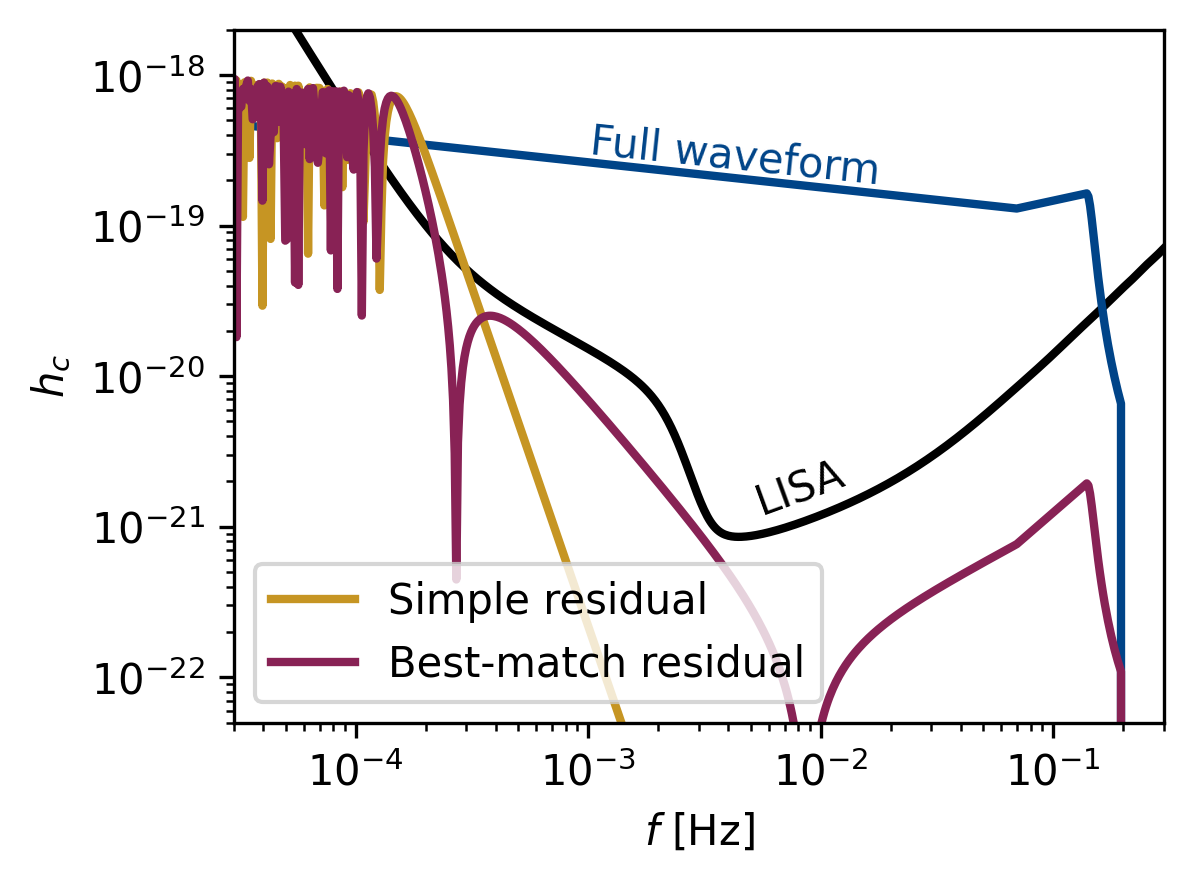}
    \caption{{Demonstration of the residuals arising from subtracting the vacuum waveform with identical parameters $h_i^{\rm vac}$ or the best match vacuum waveform $h_i^{\rm bf}$ from the full waveform with gas effects $h_i$. The parameters are chosen to showcase unrealistically strong EEs, to aid visualisation.}}
    \label{fig:individual}
\end{figure}
\begin{figure}
    \centering
    \includegraphics[width=1\linewidth]{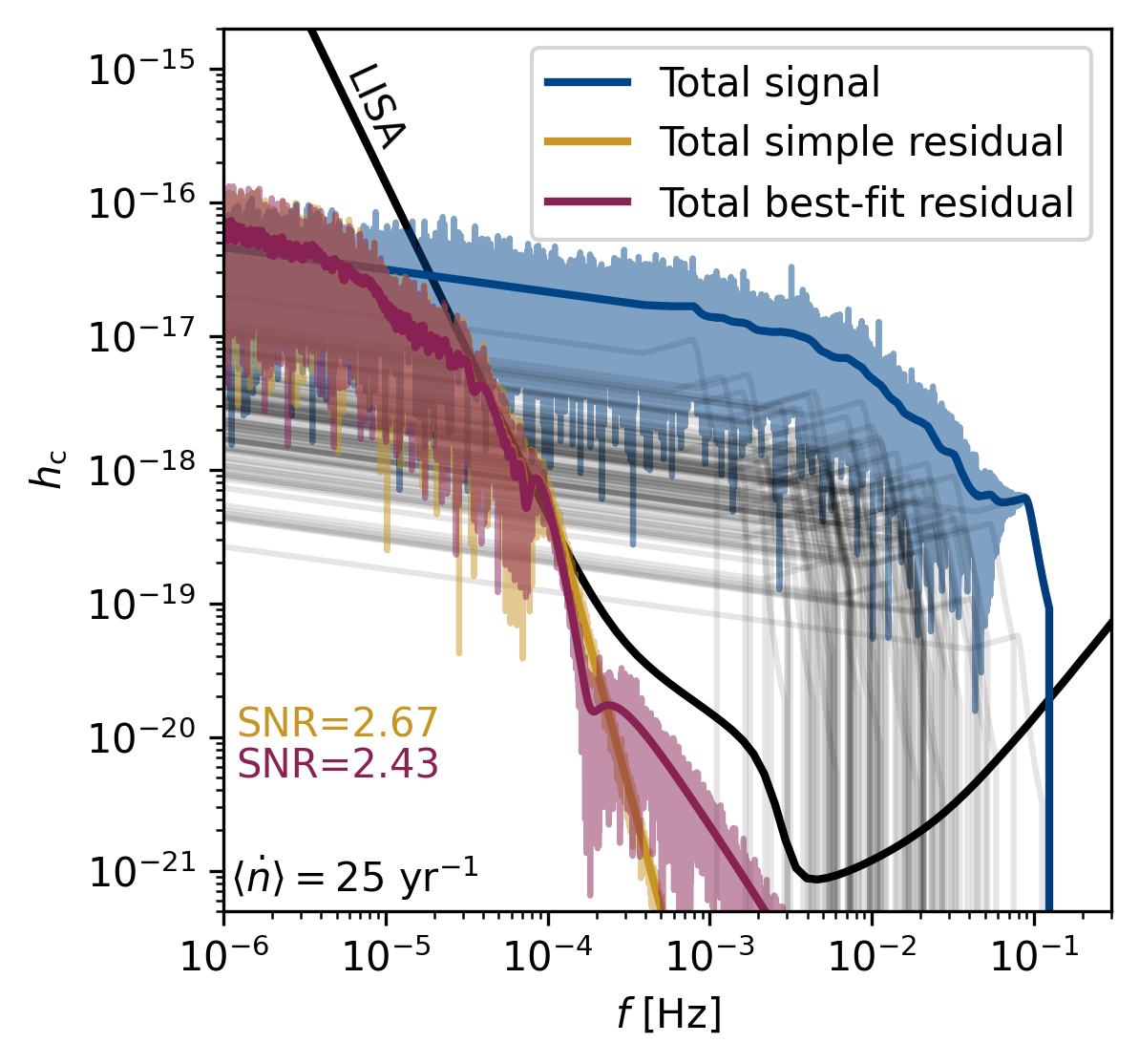}
    \caption{{Realisation of the MBH binary signals observed over 4 years in LISA. Here we consider $\sim100$ sources (thin black lines). Their total power results from the addition of individual GWs with differing phases (shown are the actual sum and the sum in quadrature, blue lines). The residuals computed with the simple or best-match method (ochre and red, here for $f_{\rm Edd}=1$) do not differ substantially in terms of their overall power distribution. This is due to the strong frequency scaling of the dephasing prescription (Eq. \ref{eq:dephasing}) being hard to mask with variations in the vacuum parameters. Overall, the residuals caused by subtracting vacuum waveforms from the signals with gas effects resemble a stochastic background. Its properties are characterized in section \ref{sec:Results}.}}
    \label{fig:example_pop}
\end{figure}
We employ waveforms of the PhenomA family \citep{ajith}, supplemented with a leading order Newtonian phase \citep{peters1964,2015moore,2018maggiore}. This is appropriate for computing residuals caused by EEs, as they primarily affect the early inspiral of binaries where relativistic effects are weak \citep{blanchet2014,2015moore}. In {the early inspiral}, the frequency-domain waveform {in the stationary phase approximation (SPA)} is given by \citep{1994cutler}:
\begin{align}
    \tilde{h}(f) &= \frac{1}{D_L}\,\sqrt{\frac{5}{24}}\,\pi^{-2/3}\,\left(\frac{G\mathcal{M}}{c^3}\right)^{5/6}\,f^{-7/6}\,e^{i\Psi(f)}, \\
    \label{eq:vac_phase}
    \Psi(f) &= 2\pi f t_{\rm c} - \phi_{\rm c} - \frac{\pi}{4} + \frac{3}{128}
    \left( \frac{\pi G \mathcal{M} f}{c^3} \right)^{-5/3},
\end{align}
where $f$ is the detector-frame frequency, $D_L$ is the luminosity distance, $\mathcal{M}$ is the binary chirp mass, and $t_{\rm c}$ and $\phi_{\rm c}$ are the coalescence time and phase, respectively. The amplitude is sky-averaged as customary.

To model GW emission from a population of binaries over a $4$-year LISA observation, we draw the total number of sources from a Poisson distribution with mean $4\,{\rm yr}\times\langle \dot{n} \rangle$. For each source, the waveform parameters are then drawn from the population distributions described in section \ref{sec:Methods:pop}, with an initial orbital phase uniformly sampled in $[0,2\pi)$ and a coalescence time uniformly distributed between $0$ and $4$ yr\footnote{This choice is mostly arbitrary. It neglects to model ``boundary effects" for a small subset of systems, such as binaries that enter the LISA band but do not merge before the end of the mission, or sources that only partially overlap. We expect the effect of this to only marginally influence our conclusions, as discussed further in section \ref{sec:conclusion}.}. EEs due to gas accretion are incorporated through a dephasing prescription {arising from Eq. \ref{eq:gaschirp}} that modifies the GW phase relative to the vacuum evolution:
\begin{align}
    \Psi \rightarrow \Psi_{\rm GW} + \delta \Psi.
\end{align}
As customary {in the SPA}, the dephasing is computed using the approximation \citep{kocsis,pedo}:
\begin{align}
    \frac{{\rm d}^2\delta\Psi}{{\rm d}f^2} \approx-2\pi\,\frac{2\,\dot{\Omega}_{\rm gas}}{\dot{f}_{\rm GW}^2},
\end{align}
and is defined such that $\delta\Psi$ vanishes at the time of merger. {This results in the following dephasing prescription:}
\begin{align}
\label{eq:dephasing}
    \delta\Psi=
-
\frac{25 A f_{\rm Edd}}{26624 \pi^{13/3}\,}\,
\frac{1+q}{q}
\frac{m_p\, c^9}
{\sigma_T\, G^{7/3}\, \mathcal{M}^{10/3}}\,
f^{-13/3},
\end{align}
{where we used the seminal result of \cite{peters1964} for $\dot{f}_{\rm GW}$.}
{Note that this treatment of EEs is valid as long as the gas effects are small with respect to the GW driven evolution, and as long as the usual SPA requirements are met\footnote{{In our computations, we verify these conditions with a simple numerical check. They are never violated for any realisation of sources within a frequency range $f> 10^{-5}$ Hz, and therefore never affect any relevant result in the LISA band. Below this threshold, the dephasing can become large enough to entirely saturate the signal. In such a case the residual becomes comparable to the overall waveform, requiring entirely different modeling.}} \citep[see][]{1994cutler,2018maggiore}. Note also that Eq. \ref{eq:dephasing} shows that dephasing due to accretion will typically be significantly stronger for lighter sources due to the strong scaling with chirp mass, as discussed in e.g. \cite{garg2022,2023zwick,2025zwick}.} For a given choice of the Eddington ratio and coefficient $A$, the dephasing is fully specified. Throughout this work we adopt $f_{\rm Edd}=1$ and $A=3$ as {fiducial choices. From Eq. \ref{eq:dephasing} we see that the dephasing scales linearly with these parameters, resulting in a linear behavior for the residual of an individual waveform (see Figure \ref{fig:individual}). In turn, this will result in the overall residual scaling as the square root, due to quadrature.} Finally, note that the contribution to the GW phase due to gas accretion must be evaluated using the relation $f = 2\Omega/(1+z)$.

To construct the residuals associated with imperfect subtraction, we compute both the environmentally affected waveform $h_i$ and the corresponding vacuum waveform $h_i^{\rm vac}$ for each source $i$. {We then employ two separate methods to estimate the residuals: In the ``simple" method,} the total residual is obtained as the superposition of the difference between full waveforms and vacuum waveforms:
\begin{align}
\label{eq:diff}
    \tilde h_{\rm tot}^{\rm res} = \sum_i  \tilde h_i - \tilde h_i^{\rm vac}.
\end{align}
Note that this simple addition of waveform differences will be incoherent due to the effectively random phases. We model detector noise by using the LISA PSD from \citep{2019robson}, and the signal-to-noise ratio (SNR) of a GW $h$ is computed via the noise-weighted inner product \citep{2018maggiore}:
\begin{align}
    {\rm SNR}^2(h) = 4 \int_0^\infty \frac{|\tilde{h}(f)|^2}{S_n(f)}\,{\rm d}f.
\end{align}
This procedure to construct an overall residual only approximates the outcome of a full global-fit inference, {which is beyond the scope of this work}. {As a compromise between a full inference and the simple method, we also perform waveform subtraction where the basic unperturbed vacuum waveform is replaced by the best-matching vacuum waveform, $h_i^{\rm bf}$. The best-match waveform \citep[see e.g.][]{1992finn,1996owen,2018maggiore} is found by maximising the inner-product $(a_i,h_i)$ of a waveform $a_i$ with respect to the true signal over the vacuum parameters. It corresponds to the maximum likelihood waveform (the outcome of a full inference) in the limit of high SNR, appropriate priors and waveform model, as well as Gaussian stationary noise. In practice, we compute the best match parameters over the parameters $[t_{\rm c}, \phi_{\rm c},z,\mathcal{M},\mu]$ of the PhenomA waveforms ($\mu$ being the reduced mass) with a \texttt{scipy} minimization routine, initializing the search in the vicinity of the vacuum parameters. Then, the residual is computed as in Eq. \ref{eq:diff} with $h_i^{\rm vac} \to h_i^{\rm bf}$.}

{Figure \ref{fig:individual} shows an example of the signal of an individual MBH binary source, as well as the residuals computed with the simple and best-match methods. As we can see, the best-match residual decays slightly faster than the simple residual, with additional features emerging at higher frequencies below the LISA sensitivity. However, the difference is not substantial in terms of the overall power distribution above the LISA PSD. This can be explained by the fact that gas induced EEs of the form given by Eq. \ref{eq:dephasing} scale extremely strongly with frequency and are therefore hard to mask with variations in vacuum parameters. In fact, a scaling of $f^{-13/3}$ corresponds to a ``negative 4PN" contribution to the GW phase in the standard post-Newtonian (PN) classification \citep[see Refs.][]{2014barausse,2020cardoso}, and is almost orthogonal to other PN contributions to the phase}. Fig. \ref{fig:example_pop} shows instead an example of the summed GW signal and GW residual from a draw of 100 sources, corresponding to a mean merger rate of 25 yr$^{-1}$. We can see that a realization for these fiducial parameters likely produces a residual above LISA sensitivity.  {We also see that the best-match and simple residuals differ little in terms of overall power distribution for a realistic population of sources. Given this fact, we will employ the latter for the rest of this work due to the factor $\sim 100$ faster evaluation.}  We discuss the limitations of these approaches further in section \ref{sec:conclusion}.

\section{Results}
\label{sec:Results}
\subsection{Characterization of the residuals}
\label{sec:Results:examples}
\begin{figure}
    \centering
    \includegraphics[width=1\linewidth]{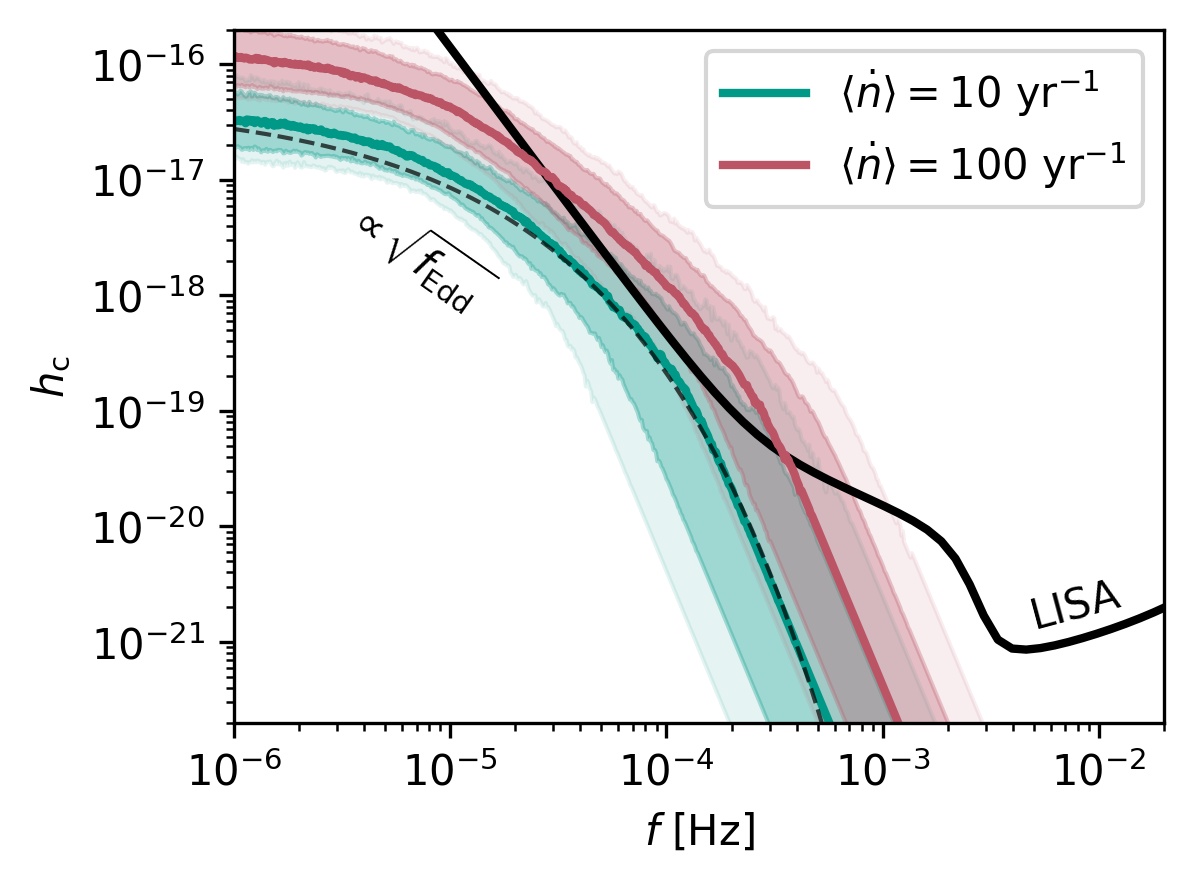}
    \includegraphics[width=1\linewidth]{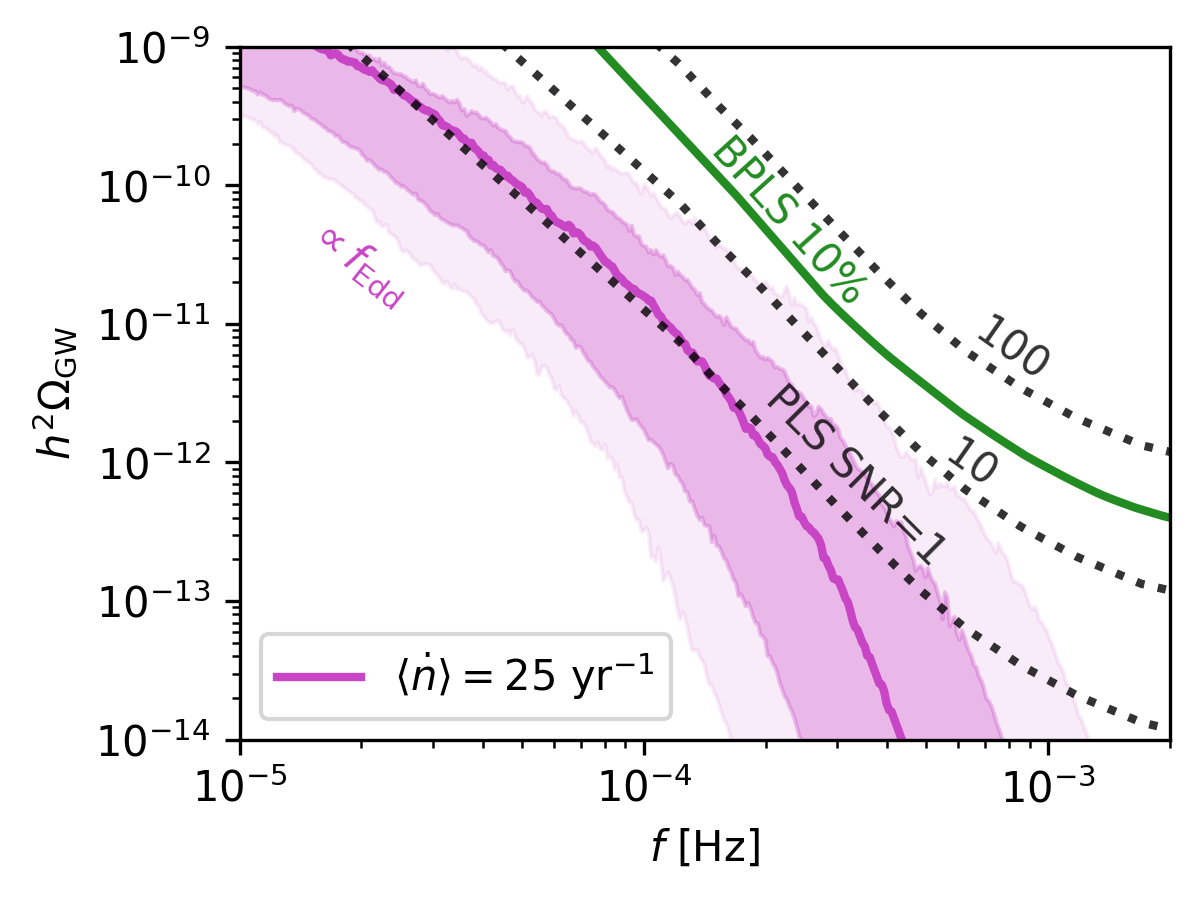}
    \caption{Top panel: Residuals calculated from $10^3$ draws from the representative population distributions discussed in section \ref{sec:Methods:pop}. The medians (solid) are bracketed by the 68th and 90th percentiles (shaded). They can be fit by the simple function shown in Eq. \ref{eq:fit1} (dashed line, here for $\langle \dot n \rangle=10$ yr$^{-1}$). {Bottom panel: GW energy density of the residuals for $\langle \dot n \rangle=25$ yr$^{-1}$ compared to the PLS curves with different SNR thresholds (dotted) and BPLS curve (green solid line)  for a 10\% error in the noise covariance matrix (see section \ref{sec:Results:detectability}).}}
    \label{fig:stat}
\end{figure}

\begin{figure*}
    \includegraphics[width=1\linewidth]{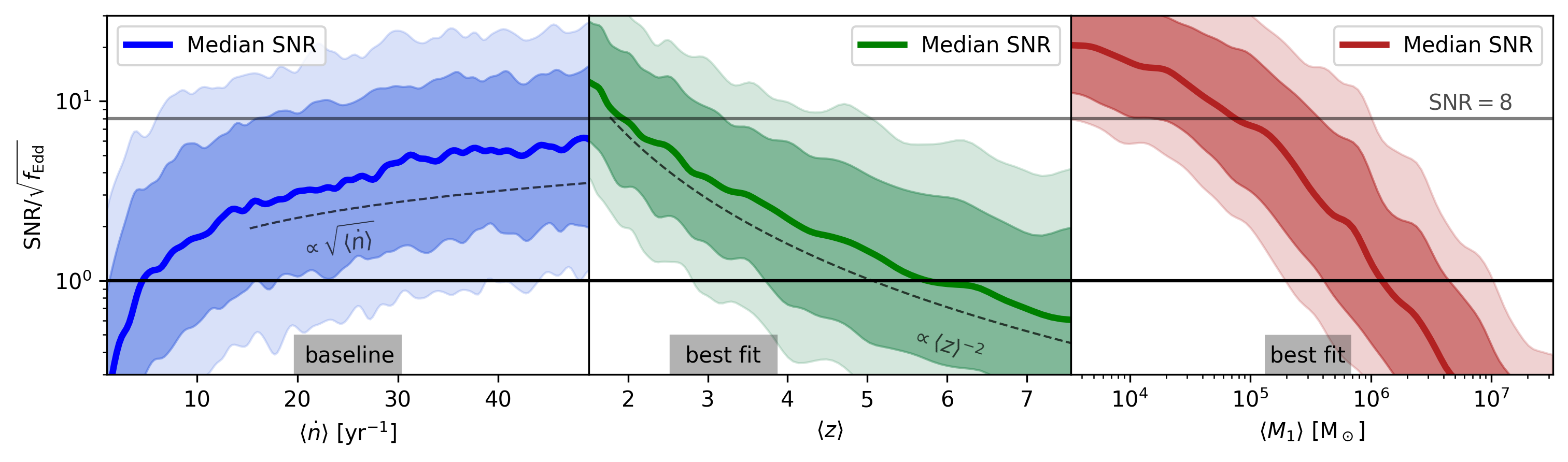}
    \caption{SNR of the residual background as a function of the population hyper-parameters. The median (solid) and percentiles (shaded) are calculated from 300 draws for each value of the mean merger rate $\langle \dot n \rangle$ (blue), mean redshift $\langle z \rangle$ (green) and mean primary mass $\langle M_1 \rangle$ (red). Shown are also the hyper-parameter ranges discussed in section \ref{sec:Methods:pop} (grey areas).}
    \label{fig:dependence}
\end{figure*}

Fig. \ref{fig:stat} illustrates the statistical properties of the overall residual obtained from $10^3$ independent realizations. In each realization, MBH binary sources are drawn from the population distributions described in section \ref{sec:Methods:pop}, with population hyper-parameters randomly chosen from the ranges specified in Eqs. \ref{eq:popparam1} to \ref{eq:popparam2}. The uncertainties in the underlying source population translate into a spread in the predicted residual background. This results in the 68\% percentile corresponding approximately to a factor of three variation in amplitude and a factor of two variation in frequency about the median realization. At frequencies below a few $10^{-6}$~Hz, the amplitude of the median residual approaches that of the quadrature sum of all MBH binary signals, reflecting the fact that the GW phases of essentially all systems are significantly modified by gas accretion in this regime. At higher frequencies, the residual follows an exponential decay. For merger rates $\langle \dot n \rangle \, \gtrsim 15~{\rm yr}^{-1}$, the median residual exceeds LISA sensitivity. Higher merger rates give rise to a PSD bump similar to the Galactic white dwarf binary foreground, though shifted to frequencies around $\sim 10^{-4}$ Hz.

We propose a simple fit that characterizes the median residual, as a function of the expected binary merger rate, the typical Eddington fraction of LISA MBH binaries, and assuming the population distributions discussed in section \ref{sec:Methods:pop}:
\begin{align}
\label{eq:fit1}
    h_{\rm c}^{\rm res} \approx C \exp\left[ -\left(\frac{f}{f_{\rm c}} \right)^{1/2}\right],
\end{align}
where:
\begin{align}
    C&= 6.7\times 10^{-17} \left(\frac{\langle \dot n \rangle}{20\,{\rm yr}^{-1}}\right)^{1/2}f_{\rm Edd}^{1/2}\\
    f_{\rm c} &= 3.9\times 10^{-6} \left(\frac{\langle \dot n \rangle}{20\,{\rm yr}^{-1}}\right)^{1/6} \, {\rm Hz}.
\end{align}
{Here, the square root scaling in $f_{\rm Edd}$ and $\langle \dot n \rangle$ simply come from the addition of individual signals in quadrature. The scaling of $f_{\rm c}$ and the overall exponent instead arise from combination of the vacuum GW power law of $f^{-1/6}$, the $f^{-13/3}$ scaling of the gas dephasing prescription and the overall distribution of binary sources.} The fit performs adequately (remaining within the 68\% percentile of realizations) over the key frequency range of $10^{-6}$ to $10^{-3}$ Hz for all merger rates $\langle \dot n \rangle \, \gtrsim 5~{\rm yr}^{-1}$. For lower merger rates, the residual is instead dominated by individual instances of poorly subtracted MBH binary signals. An illustrative example of the fit function is also shown in Fig. \ref{fig:stat}. {We note that this form for the fit function is appropriate for any dephasing prescription of order ``-4PN", which include important effects such as line-of-sight acceleration \citep{2017meiron,2024MNRAS.527.8586T,kai2024}, provided that the magnitude of $C$ be adjusted accordingly.}

\subsection{Dependence on population distribution parameters}
\label{sec:Results:SNR} 

We now characterize the dependence of the residual background on the hyper-parameters that define the underlying MBH binary population. To this end, we vary the mean of one hyper-parameter $x$ while sampling all others from their fiducial ranges. For the selected hyperparameter, we generate $300$ independent population realizations at fixed values of $\langle x\rangle$, and compute the corresponding SNRs statistics of the residual. {Here, we use the SNR as a simple quantity to express excess power over the LISA PSD, noting that residuals with SNR $\gtrsim1$ can bias the parameter estimation of other sources \citep[see e.g., Refs.][]{2007cutlervallis,2021antonelli}. Wether the overall residual background is detectable and distinguishable by itself requires further investigation \citep[see e.g., Ref.][for an interesting approach]{Dideron:2022tap}, and is only partially addressed in this work in section \ref{sec:Results:detectability}.}

Fig. \ref{fig:dependence} illustrates this procedure for variations of the expected mean merger rate, the mean merger redshift, and the mean primary black hole mass. As expected from quadrature, the median SNR of the residual increases with the square root of $\langle \dot n \rangle$, reaching typical values of ${\rm SNR} \sim 3$ for a fiducial choice of $\langle \dot n \rangle=20$ yr$^{-1}$ {with $f_{\rm Edd}=1$, and scaling with $\sqrt{f_{\rm Edd}}$}. The dependence on the mean redshift of mergers is also described by a power law, with ${\rm SNR} \propto \langle z \rangle^{-2}$. In contrast, the dependence on the primary mass exhibits an approximately exponential suppression.

The value of the SNR, marginalized over the population hyper-parameters defined in Eqs. \ref{eq:popparam1} to \ref{eq:popparam2}, is:
\begin{align}
\label{eq:snr_final}
    {\rm SNR} =  3.2 ^{+5.4}_{- 1.9}\times \sqrt{f_{\rm Edd}\, \langle \dot n \rangle /(20\, {\rm yr}^{-1})}\,,
\end{align}
where the uncertainties correspond to the 68th percentile. {We note again that here $f_{\rm Edd}$ is interpreted as the typical Eddington ratio for LISA MBH binaries. Alternatively, $\langle \dot n \rangle$ can be interpreted as the merger rate of sources that have an Eddington ratio of $f_{\rm Edd}$ or above \citep[see e.g. the thresholds in][]{2022mangiagli}, and Eq. \ref{eq:snr_final} becomes a lower bound. Aspects that can significantly alter the results from this baseline estimate will be discussed in section \ref{sec:conclusion}.}

\subsection{Distinguishability of residuals as a stochastic background}
\label{sec:Results:detectability}
{While pollution from the residual background can bias the reconstruction of individual sources, a direct detection would provide valuable information about the typical accretion rates of MBH binaries. Estimating the sensitivity of a GW detector to a stochastic background requires more than simple SNR calculations \citep[see e.g.][]{2010adams,ThraneRomano:2013}. Here, we assess the prospects for identifying the residuals computed above as a stochastic signal by comparing their amplitude to recently derived Bayesian power-law sensitivity (BPLS) curves \citep[see][]{2025pozzoli}. The BPLS depends both on the required log-Bayes factor and on the assumed uncertainty in the detector noise covariance matrix, the latter being a critical factor in distinguishing stochastic signals \citep{2024muratore}. Following \cite{2025pozzoli}, we adopt a log-Bayes factor of 1, corresponding to strong evidence, and consider an optimistic representative noise uncertainty of 10\%. This constitutes a more stringent criterion for detectability and distinguishability than simpler power-law sensitivity (PLS) curves \citep{ThraneRomano:2013}, which are instead defined solely in terms of a required SNR threshold.
To perform the comparison, we compute the effective energy density of the residual background as:
}
\begin{align}
\Omega_{\rm GW}^{\rm res}(f) = \frac{4 \pi^2 f^3}{3 H_0^2 \, T_{\rm obs}} \, 2\langle |\tilde{h}^{\rm res}_{\rm tot}(f)|^2 \rangle^{\rm avg},
\end{align}
{where $T_{\rm obs}=4$ yr is the observation time, $\langle \cdot \rangle^{\rm avg}$ denotes the ensemble average and the factor 2 normalizes to a one-sided PSD \citep{2019robson}.}

{The bottom panel of Fig.~\ref{fig:stat} shows the resulting $\Omega_{\rm res}(f)$ for a merger rate of $\langle \dot n \rangle = 25\,\mathrm{yr}^{-1}$, together with selected BPLS and PLS curves interpolated from Ref.~\cite{2025pozzoli}. The median residual energy density slightly exceeds the PLS corresponding to an SNR threshold of 1, consistent with Eq. \ref{eq:snr_final}. This confirms that the residuals identified in this work are likely large enough to bias the inference of other signals. In contrast, the residual spectrum lies well below the BPLS curve corresponding to a log-Bayes factor of 1 (strong evidence), even under the most optimistic assumption of a 10\% fractional uncertainty in the noise covariance matrix. In fact, only the loudest $\sim 5\%$ of realizations reach this detection threshold. Overall, these results indicate that the residual is unlikely to be confidently distinguishable from instrumental noise, although it is sufficiently strong to affect parameter estimation in the context of the global fit.
}

\section{Discussion and conclusions}
\label{sec:conclusion}
{As follows from Eq. \ref{eq:dephasing}}, the dominant contribution to the residual arises from the lighter end of the MBH binary population, which is more susceptible to EEs and therefore more strongly mismodeled by vacuum templates. This part of the MBH population is uncertain, due to the stronger sensitivity to merger delays and seeding prescriptions \citep{2022lisaastro}, and may even include contributions from exotic formation channels such as primordial black holes \citep{deluca2021,2022yang}. {Crucially, our analysis assumes no correlation between the binary parameters and the expected value of $f_{\rm Edd}$. This simplifying assumption allows $f_{\rm Edd}$ to be interpreted as a representative value for a population characterized by a merger rate $\langle\dot n\rangle$. In reality, we expect a correlation to exist, and there are indications that lighter MBH binaries may require more sustained gas accretion to overcome the final-parsec barrier and reach the LISA band, implying systematically higher Eddington ratios in this mass range \citep[see Ref.][]{2022mangiagli}. A more realistic treatment would therefore model $f_{\rm Edd}$ as a function of mass, for example by linking the waveform subtraction directly to a semi-analytical model for merger rates and accretion histories. This would not only refine the characterization of the residual background, but also clarify the relative contribution of different mass scales. Beyond the characterization of the residual, the dominance from lighter mass MBH binaries} suggests that targeted mitigation strategies may be optimal, for example by identifying and reanalyzing subsets of sources with waveform models that explicitly incorporate EEs, while treating heavier systems with standard vacuum templates \citep[see e.g., Refs.][]{garg2022,2023zwick}. An advantage of this approach is that it can be implemented a-posteriori, once the lighter MBH signals have already been identified using vacuum templates.

Another approach is suggested by a limitation of our methodology, i.e. that the analysis is performed entirely in the frequency domain, where residuals are effectively treated as a stochastic background. In the time domain it may be possible to correlate residual features with individual merger events, thereby enabling the identification and excision of contaminated segments of the data stream. This is also necessary to properly account for boundary effects associated with signals that only partially overlap, or that do not merge within the LISA mission lifetime. Typical LISA binaries from the population distributions described in section \ref{sec:Methods:pop} enter the band at $\lesssim 10^{-4}$ Hz with chirp masses $\sim 10^5$ M$_\odot$ at $z \sim 2$, corresponding to inspiral timescales of $T_{\rm in}\sim 1$ yr. The average number of sources overlapping at any given time is therefore $\langle\dot n \rangle\, T_{\rm in} $, or about one quarter of the total number of sources expected within the LISA lifetime. Accounting for this reduced number of overlapping signals by performing the analysis in the time domain would, in principle, reduce the effective SNR of the residual. However, it would do so by at most a factor of two, from quadrature. Most likely, the reduction would in fact be even smaller, since the dominant contribution to the residual arises from lighter, longer-lived sources. A full quantification of this effect that accounts for the distribution of inspiral timescales is left for future work.

In this work individual residuals are computed by directly subtracting waveforms with identical vacuum parameters, rather than performing a full parameter inference for each source. {We have verified that replacing this simple subtraction with best-match vacuum waveforms does not significantly change our conclusions, owing to the steep frequency dependence of the dephasing prescription, which limits the ability of vacuum parameters to absorb the EEs. While this provides a reasonable and computationally efficient approximation, it does not substitute for a full Bayesian analysis performed within the global-fit framework. In general, the problem of treating waveform systematics and residuals in a Bayesian framework is very complex and several approaches have been proposed \citep[see Refs.][]{2014moore_margi,2015gair}. We highlight in particular Ref. \cite{owen23} where waveform systematics are treated by performing a marginalisation over the yet unknown higher order coefficients of the PN series for the GW phase. Interestingly, such a framework could be naturally extended to incorporate EEs phenomenologically, i.e. by including additional negative-PN contributions, most notably the $-4$PN term. Note here that residuals due to mismodelled EEs will affect the low frequency part of the LISA band, while the contrary is true for systematics due to unmodelled higher order PN terms.} 

Finally, we comment on our modeling of binaries and their environments. In this work, we have focused exclusively on phase modifications induced by gas accretion, as gas-assisted binary formation is considered a robust channel for LISA sources. Other formation pathways, such as MBH triples \citep{2019bonetti}, may also produce similar EEs. We have also neglected the impact of orbital eccentricity, which is known to be a characteristic signature of binaries evolving in gas \citep{2009cuadra,Roedig_Trqs+2012} and may be directly measurable in their GW signals \citep{2024garg}. Residual eccentricity in the LISA band is expected to enhance the importance of EEs \citep{2020moore,pedo,2025zwickecc_pop}, introduce additional waveform complexity, and further exacerbate mismodeling and parameter degeneracies. Including it is therefore likely to increase both the amplitude and the complexity of the residual background.
\newline

With these discussion points in mind, our analysis shows that that imperfect subtraction of GW signals using vacuum templates will very likely give rise to a residual background above the sensitivity of LISA. This highlights an additional aspect of EEs and their consequences for the analysis of future data from space based detectors: Even in a scenario in which they are undetectable in individual signals, the cumulative effect of EEs on a population level {can play a crucial role} in the context of a global fit.

\section*{Data availability}
The code used to produce the results will be shared upon reasonable request.
\section*{Acknowledgments}
L.Z. is supported by the European Union’s Horizon 2024 research and innovation program under the Marie Sklodowska-Curie grant agreement No. 101208914. L.Z. acknowledges Dr. Arianna Renzini and Prof. Lucio Mayer for useful discussion.

\bibliography{residuals}


\end{document}